\documentstyle[epsfig]{mn2e}

\def\simgt{\mathrel{\lower0.6ex\hbox{$\buildrel {\textstyle >}
 \over {\scriptstyle \sim}$}}}
\def\simlt{\mathrel{\lower0.6ex\hbox{$\buildrel {\textstyle <}
 \over {\scriptstyle \sim}$}}}

\newcommand{\ltsim}{\mbox{{\raisebox{-0.4ex}{$\stackrel{<}{{\scriptstyle\sim}}
$}}}}
\newcommand{\mc}{\multicolumn}
\def \kband{$K$-band }
\def \USS{74~MHz USS\,}

\begin{document}

\title[$K-$band imaging of a sample of Ultra-Steep-Spectrum radio
  sources selected at 74~MHz]{Near-infrared $K-$band imaging of a sample
 of Ultra-Steep-Spectrum radio sources selected at 74~MHz}

\author[Matt J.~Jarvis et al.]
{Matt J.~Jarvis$^{1}$\thanks{Email: mjj@astro.ox.ac.uk}, Maria J.~Cruz$^{1}$,
  Aaron S.~Cohen$^{2}$, 
\and
Huub J.A.~R\"ottgering$^{3}$, Namir E.~Kassim$^{2}$ \\
\footnotesize
\\
$^{1}$Astrophysics, Department of Physics, Keble Road, Oxford, OX1 3RH, UK \\
$^{2}$Naval Research Laboratory, Code 7213, Washington, DC, 20375 USA \\
$^{3}$Leiden University, Sterrewacht, Oort Gebouw, P.O. Box 9513, 2300 RA Leiden, The Netherlands
}

\maketitle

\begin{abstract}
In this paper we present near-infrared $K-$band imaging of a sample of
ultra-steep spectrum (USS) radio sources selected at 74~MHz. The dual
selection criteria of low frequency and USS means that we should be
sensitive to the highest redshift ($z > 5$) radio galaxies. We have obtained
$K-$band magnitudes for all of the objects in our sample of 26 and
discuss the properties of each.

There is a pronounced bias in this sample towards fainter magnitudes
and thus higher redshifts when compared to complete unfiltered samples
such as 7CRS (Willott et al. 2002), implying that the steep-spectrum
technique is still viable at 74~MHz. However, there are more bright
($K < 17$) sources in the 74~MHz sample than in a similar sample
selected at 151~MHz, namely 6C*. This is principally due to the
additional selection criterion of a small angular size for the radio
sources in 6C*, four of the six sources in the 74~MHz USS sample with
$K < 17$ have angular sizes $> 15$~arcsec (the angular size cutoff of
6C*).

We find that the distribution of \kband magnitudes from a sample
selected at 74~MHz is statistically indistinguishable from the 6C*
sample, when similar angular size filtering is applied to the 74~MHz
sample. 

\end{abstract}

\section{INTRODUCTION}

The search for high-redshift radio galaxies has been somewhat overshadowed in
recent years by the discovery of significant numbers of radio-quiet
quasars from the Sloan Digital Sky Survey (e.g. Fan et al. 2003). However, the
bright point-source nature of the quasars means it is extremely
difficult to study the stellar populations of the host galaxies of
these powerful active galactic nuclei (AGN). For this reason alone it is still useful to
search for distant radio galaxies.

Furthermore, there are extremely important roles that radio
galaxies may play in our understanding of general galaxy formation and
evolution. First, the fact that they are selected via radio emission means
that they are essentially free of selection effects inherent to
optical surveys, such as dust obscuration, which may be more prevalent
in the early Universe (e.g. Archibald et al. 2001), and particularly
around massive galaxies (Stevens et al. 2004). Second, in radio
galaxies the powerful nuclear emission is obscured by the `dusty
torus' in orientation-based unification schemes, thus we have an uninhibited view
of any neutral gas all the way up to the host galaxy, without the
problem of the ionised Stromgren sphere along the line-of-sight which
quasars suffer from.
This means that the gas, which possibly helped in the formation of the
galaxy and in fuelling the AGN, can be probed up to the
vicinity of the host galaxy itself (e.g. Jarvis et al. 2003; Wilman et al. 2004).
Third, if we discover powerful radio galaxies within the epoch of
reionization we would be able to be able to probe the 21~cm forest at
this important epoch in the history of the Universe (e.g. Carilli,
Gnedin \& Owen 2002). Fourth, it is also now known that powerful
radio galaxies in the high-redshift Universe trace overdensities of
galaxies (e.g. Venemans et al. 2002; Miley et al. 2004) which may go on to become the richest clusters in the nearby
Universe and thus provide a probe of large-scale structure formation
over all cosmic epochs.
A fifth reason is that they are very luminous allowing for detailed studies 
of cold, warm and hot gas using millimetre, optical and x-ray
telescopes (e.g. Stevens et al. 2004; Villar-Mart\'in et al. 2003;
Carilli 2003). 


In the past decade the search for radio galaxies at the highest
redshifts has been optimized using various selection techniques. The
most widely used of these is that of a steep-radio-spectral
index. This works because radio spectra are generally concave, thus
they steepen at high frequencies, this is particular true at the
highest redshifts where the inverse Compton losses, due to the
increase in temperature of the cosmic microwave background, are higher. 
 Further, the observed rest frame frequencies are higher for 
high-redshift sources.  Therefore, by filtering a sample to include just 
those sources with steep spectral indices one would expect to select in 
favour of high redshift objects.  This idea has in fact been verified
by several groups (e.g. Blundell et al. 1998; De Breuck et
al. 2000, 2001; Jarvis et al. 2001a,b).

Another important radio selection criteria is that of low
frequency. This is because the spectra of the highest redshift sources
are redshifted towards low frequency, thus the bulk of the flux from
the source itself, although emitted at higher frequencies is observed
at much lower frequencies, and this effect obviously increases with
redshift. Thus the majority of surveys for high redshift radio
galaxies have taken place at frequencies $\ll 1$~GHz, the two most
prominent being 325~MHz (De Breuck et al. 2000, 2001) and 151~MHz
(Blundell et al. 1998; Jarvis et al. 2001a,b), which have found the
highest redshift radio galaxies to date (Rawlings et al. 1996; van
Breugel et al. 1999). Low-frequency selection also preferentially
selects the optically thin lobe emission, which is independent of
orientation, unlike high-frequency selection which preferentially
selects the optically thick emission from the core of the AGN, which
means that low-frequency surveys are crucial for unification studies
(e.g. Willott et al. 2000).  The next step in this type of survey
would be to push the observed frequency lower.


The new 74~MHz system on the Very Large Array (VLA), fully implemented in 1998 
(Kassim et al. 1993), has opened up a new window into the previously 
unexplored regime of very low-frequency radio observations at 
high sensitivity and sub-arcminute resolution.  Therefore, we have
initiated a survey at this low frequency, with steep-spectral index
selection, to continue the search for the highest-redshift radio galaxies.
The radio data for this survey are described in Cohen et al. (2004),
and has $\sim 10$~times the resolution and sensitivity of 8C.
In this paper we present our near-infrared $K-$band follow-up
observations of all the steep-spectrum sources in this sample.

The paper is set out as follows, in section~\ref{sec:sample} we summarize
how the sample was selected and in section~\ref{sec:observations} we describe the $K-$band observations which are
then presented in section~\ref{sec:results}. In section~\ref{sec:analysis} we
compare the distribution of $K-$band magnitude with similar samples
selected at higher frequencies and in section~\ref{sec:conclusions} we
discuss our results.

\section{Sample Selection}\label{sec:sample}

Full details of how the sample was selected can be found in Cohen et
al. (2004). For clarity we summarise the selection criteria here.

\begin{itemize}
\item $S_{74 \rm MHz} \geq 0.1$~Jy. This is a lower limit as the
  sensitivity is a function of
  position on the sky, due to areas of higher noise. We refer the
  reader to Cohen et
  al. (2004) for full details.

\item Spectral index between 74~MHz and 1400~MHz $\alpha_{74}^{1400} <
  -1.2$, where $S_{\nu} \propto \nu^{\alpha}$. The NRAO VLA Sky Survey
  (NVSS; Condon et al. 1998) was used to measure the 1400~MHz flux
  densities as the low-angular resolution ($\sim 40$~arcsec) is well
  matched to that of the 74~MHz observations ($\sim 25$~arcsec),
  therefore we do not expect any extended structure to be resolved out
  in either survey.

\end{itemize}

This leads to a sample of 26 sources detected over a total area of
0.05~sr.

\section{\kband imaging}\label{sec:observations}
\subsection{Observations}

The \kband (2.2$\mu$m) imaging of the \USS sample was made over several
flexibly scheduled runs at the United Kingdom Infrared Telescope
(UKIRT), beginning in Dec 2003 up to March 2004. 
The observations were made using the high-resolution infrared camera on UKIRT, UFTI
(UKIRT Fast-Track Imager), which comprises a $1024 \times 1024$ HgCdTe
array, with a plate scale of 0.091 arcsec pixel$^{-1}$, giving a field
of view of $92 \times 92$ arcsec$^{2}$. All observations were made in
photometric conditions, with seeing always $< 0.6$~arcsec, with the
tip-tilt system. 

In order to subtract the rapidly changing sky background at these
wavelengths, to provide a good flat-field and to minimise the effects
of cosmic ray contamination and bad pixels, we used the standard
observing strategy of offsetting the position of the telescope by
$\sim 10$\,arcsec between each exposure. The offsets were arranged in
a $3 \times 3$ mosaic of nine exposures. The integration time varies
between objects in the sample, with the fainter sources being observed
for longer.
A summary of all the \kband observations is given in
Table~\ref{tab:im_journalk}.

\subsection{Data reduction}\label{sec:datareduction}

The \kband images were reduced using standard procedures. We first
subtracted the dark current from each image. We then divided by the
normalised flat-field, created by combining the nine exposures of the
particular field with a median filter, which removes any objects which
appear in different positions on the chip over the nine exposures. To
combine the individual images we registered all of the frames using a
bright star which was present in each of the nine pointings. In the cases where there
was no bright star in the images, the offsets recorded in the image
headers were used to align the frames. The registered images were then
combined using an average clipping procedure to reject pixels more
than 4$\sigma$ away from the median of the distribution. 

Astrometry for each image was achieved by identifying sources
in the image with objects in the POSS-II. 
In all cases, apart from J1252.7+2207, we were able to identify three or more sources on our
images with objects on the finding charts, and in these cases the
astrometry, performed with the KARMA-Koords task package
(Gooch 1996; http://www.atnf.csiro.au/computing/software/karma/) which performs a
simple linear fit to the plate solution, and for UFTI this is typically accurate to
$\ltsim\, 1$ arcsec. Photometric calibration was performed using at
least two UKIRT faint standard stars on each night where data were taken.
The reduced images are shown in figure~\ref{fig:kband_images} with the
Faint Images of the Radio Sky at Twenty centimetres (FIRST; Becker,
White \& Helfand 1995) radio contours overlaid. The FIRST survey was
carried out at 1.4~GHz and has
an angular resolution of $\sim 5$~arcsec.

\begin{table*}
{\caption {\label{tab:im_journalk}Log of the $K-$band imaging observations for
  our 74~MHz USS sample. }}
\begin{center}
\begin{tabular}{lll}
\hline\hline \mc{1}{c}{Source} & \mc{1}{c}{Dates Observed} & \mc{1}{c}{Total exposure time / s} \\
\hline\hline 
J1225.0$+$2146 & 2004-02-12 & 2700 \\ 
J1226.3$+$2418 & 2004-01-19 & 540 \\ 
J1228.9$+$3114 & 2003-12-24 2004-01-17 2004-01-19 2004-04-09 & 3240 \\
J1229.1$+$3040 & 2003-12-24 2004-01-17 2004-01-19 2004-02-12 & 3240\\ 
J1229.9$+$3712 & 2003-12-24 2004-01-17 & 1080 \\ 
J1230.2$+$2326 & 2004-01-19 & 540\\
J1230.6$+$3247 & 2003-12-24 & 540\\ 
J1231.2$+$2538 & 2004-01-19 2004-02-13 & 1620\\ 
J1231.3$+$3724 & 2004-01-17 & 540\\ 
J1231.5$+$3236 & 2004-01-17 & 540 \\ 
J1232.2$+$2814 & 2004-01-17 2004-02-20 & 2160 \\ 
J1232.6$+$3157 & 2004-02-13 & 1080 \\
J1234.3$+$2605 & 2004-01-17 & 540\\ 
J1238.2$+$2613 & 2004-01-17 & 540\\ 
J1238.8$+$3559 & 2004-01-17 & 540\\ 
J1243.7$+$2830 & 2004-01-17 & 540\\ 
J1245.9$+$3320 & 2004-02-20 & 540  \\ 
J1246.4$+$2516 & 2004-01-17  2004-02-20 & 1620\\
J1248.2$+$2747 & 2004-01-17 & 540\\
J1249.0$+$3615 & 2004-01-19 & 540\\ 
J1249.7$+$3408 & 2004-02-20 2004-02-27 2004-04-07 & 5400 \\ 
J1250.4$+$2941 & 2004-01-17 & 540 \\ 
J1252.7$+$2207 & 2004-01-17 & 540\\ 
J1253.4$+$2703 & 2004-01-17 & 540\\ 
J1253.6$+$2509 & 2004-01-17 & 540\\
J1256.9$+$2811 & 2004-01-17 2004-02-12 & 2160\\
\hline\hline
\end{tabular}
\end{center}
\end{table*} 

\section{results}\label{sec:results}

In Table~\ref{tab:mags} we present the $K-$band magnitudes for all of
the sources in our \USS sample. The magnitudes have been measured in
three apertures of 3$^{\prime\prime}$, 5$^{\prime\prime}$ and
8$^{\prime\prime}$ respectively, both to allow ease of comparison with
previous work and to make sure that a magnitude is given even when
there is a nearby object which may increase the flux in one of the
larger apertures.

\begin{table*}
{\caption {\label{tab:mags}\kband magnitudes for the \USS sample in
three different angular apertures. 
nbo denotes
that the radio galaxy is too close to a nearby object to
measure the magnitude reliably. J1229.9+3712 is not shown as it is
now omitted from the sample because the 74~MHz emission is from the
lobe of a larger source (Figure~\ref{fig:dssimage}). $\dagger$ the quoted
magnitudes are for the bright galaxy to the north east, the actual ID
may not be visible in this image (see notes on this source).}}
\begin{center}
\begin{tabular}{lllll}
\hline\hline
\mc{1}{c}{Source} & \mc{1}{c}{Magnitude from} & \mc{1}{c}{Magnitude from}
& \mc{1}{c}{Magnitude from} & \mc{1}{c}{Notes} \\
\mc{1}{c}{} & \mc{1}{c}{3'' diameter} & \mc{1}{c}{5'' diameter}
& \mc{1}{c}{8'' diameter} & \mc{1}{c}{} \\
\hline\hline
J1225.0$+$2146 & 20.12 $\pm$ 0.33  & 19.70 $\pm$ 0.32 & 19.54 $\pm$ 0.32 &   \\
J1226.3$+$2418 & 16.911 $\pm$ 0.087 &  16.764 $\pm$ 0.083 & 16.608 $\pm$ 0.080 &   \\
J1228.9$+$3114$\dagger$ & 16.502 $\pm$ 0.079 & 16.276 $\pm$ 0.073 & 16.073 $\pm$0.069 &  \\
J1229.1$+$3040 & $>20.7$ & $> 20.2$ & $> 19.7$ &  \\
J1230.2$+$2326 & 14.301 $\pm$ 0.026 & 13.922 $\pm$ 0.022 & 13.641 $\pm$ 0.019 &  \\ 
J1230.6$+$3247A & 19.681 $\pm$ 0.375 & 19.675 $\pm$ 0.406 & 19.655 $\pm$ 0.502 & ID on the western-most radio source\\
J1230.6$+$3247B & 17.979 $\pm$ 0.151 & 17.911 $\pm$ 0.156 & 17.512 $\pm$ 0.140 & ID on the radio source to the NE \\
J1231.2$+$2538 & 19.187 $\pm$ 0.225 & 18.780 $\pm$ 0.195 & 18.759 $\pm$ 0.225 & \\
J1231.3$+$3724 & 18.098 $\pm$ 0.170 & 17.870 $\pm$ 0.161 & 17.814  $\pm$ 0.177 &  \\
J1231.5$+$3236 & 17.499 $\pm$ 0.128 & 17.276 $\pm$ 0.119 & nbo & \\
J1232.2$+$2814 & 19.313 $\pm$ 0.266 & 19.363 $\pm$ 0.292 & -- &  \\ 
J1232.6$+$3157N &17.512 $\pm$ 0.110 & 17.296 $\pm$ 0.101 & 17.151 $\pm$ 0.097 & north source  \\
J1232.6$+$3157S &17.376 $\pm$ 0.103 & 17.159 $\pm$ 0.094 & 16.726 $\pm$ 0.078 & south source  \\
J1234.3$+$2605 & 18.585 $\pm$ 0.218 & 17.992 $\pm$ 0.173  & 17.786 $\pm$ 0.176 &  \\
J1238.2$+$2613 & 17.543 $\pm$ 0.130 & 17.521 $\pm$ 0.134 & 17.662 $\pm$ 0.162 &  \\
J1238.8$+$3559 & 16.961 $\pm$ 0.099 & 16.724 $\pm$ 0.091 & 16.522$\pm$  0.087 &  \\
J1243.7$+$2830 & 17.255 $\pm$ 0.113 & 16.996 $\pm$ 0.103 & 16.717 $\pm$ 0.096 & merger? \\
J1245.9$+$3320 & 18.405 $\pm$ 0.178 & 18.082 $\pm$ 0.162 & 17.695 $\pm$ 0.151 &  \\
J1246.4$+$2516 & 18.552 $\pm$ 0.187 & 18.507 $\pm$ 0.192 & 18.516 $\pm$ 0.223 &  \\
J1248.2$+$2747 & 17.672 $\pm$ 0.139 & 17.508 $\pm$ 0.134 & 17.413 $\pm$ 0.143 & \\
J1249.0$+$3615 & 15.862 $\pm$ 0.054 & 15.607 $\pm$ 0.048 & 15.500 $\pm$ 0.047 & merger?  \\
J1249.7$+$3408 & 20.093 $\pm$ 0.396 & 19.576 $\pm$ 0.320 & 19.483 $\pm$ 0.340 &  \\
J1250.4$+$2941 & 16.815 $\pm$ 0.092 & nbo & nbo &  \\
J1252.7$+$2207 & 17.856 $\pm$ 0.151 & 17.624 $\pm$ 0.142 & 17.447 $\pm$ 0.145 &  \\
J1253.4$+$2703 & 16.777 $\pm$ 0.090 & 16.592 $\pm$ 0.085 & 16.511 $\pm$ 0.087 &  \\
J1253.6$+$2509 & 18.674 $\pm$ 0.230 & 18.628 $\pm$ 0.258 & 18.503 $\pm$ 0.322 &  \\
J1256.9$+$2811 & 19.466 $\pm$ 0.288 & 19.224 $\pm$ 0.272 & 18.885 $\pm$ 0.257 &  \\
\hline\hline
\end{tabular}
\end{center}
\end{table*}

\subsection{Notes on Individual Sources}\label{sec:notes}

In this section we present notes on all of the individual sources in
the \USS sample. In the cases where there is an infrared
counterpart within 1.5~arcsec (or within the positional
uncertainty of the individual astrometerised images) of the core of a
one-component radio source, or within 1.5~arcsec of the axis along a
double-lobed source, we take this to be the true indentification (ID) of
the radio source host galaxy.

{\bf J1225.0+2146} This source is double lobed with an extended \kband
ID towards the western lobe. There are two plausible candidates for
the ID which may be interacting, at $\alpha =$12:25:02.40,
$\delta=$21:46:52.6 with $K=20.12$, and $\alpha$=12:25:02.51, $\delta =$
21:46:54.5 with $K=20.20$. 
The entry in table~\ref{tab:mags} is the magnitude for the western most ID.
There is also a faint infrared source at the centre of the western
lobe ($\alpha =$12:25:02.14, $\delta=$21:46:50.6), with a \kband
magnitude $K > 21$, this is unlikely to be the ID but could be enhanced emission due
to the passage of the radio jet.

{\bf J1226.3+2418} A large radio source with a secure ID at the centre of
the brightest radio emission.

{\bf J1228.9+3114} This source has a slightly extended radio morphology
centred at $\alpha=$ 12:28:59.58, $\delta=$+31:14:57.6. There is a bright \kband counterpart to
the north-east of this centroid at $\alpha=$12:28:59.56,
$\delta=$+31:15:01.1 which may be the host galaxy. However, this is
not centred on the bright radio emission and it is
possible that a true ID centred on the radio emission may be beyond the detection limit of our
imaging observations at $K > 20.7$ in a 3~arcsec aperture (3$\sigma$).

{\bf J1229.1+3040} This source is slightly extended in the radio map. We
have a total of 3240~second integration with UKIRT on this source with no
apparent ID down to a limiting 3$\sigma$ total magnitude of $K=20.7$ in a
3~arcsec aperture.

{\bf J1229.9+3712} The radio morphology for this source is such that it
could be the lobe of a larger source. Indeed, inspection of the FIRST radio
maps does show three aligned components with a probable core at
$\alpha=$12:30:02.98, $\delta=$37:12:46.1. The full extent of this
radio source on the sky is $\sim 3$~arcmin. The FIRST radio map
overplotted on the Digitized Sky Survey $R-$band image is shown in
figure~\ref{fig:dssimage}. There is no bright ID detected in the POSS-II image at
the position of the core.

{\bf J1230.2+2326} This source has a double-lobed radio morphology with a bright \kband ID
situated between them, along the radio axis.

{\bf J1230.6+3247} An extended radio source with two \kband counterparts close to
both of the bright radio components. 
This could be two distinct sources with both of the two infrared objects hosting an
AGN (marked A and B in figure~\ref{fig:kband_images}). We note that the peak centroid of the 74~MHz radio map is centred
closer to the south-western component with the fainter \kband ID at
$\alpha=$12:30:37.96, $\delta=$+32:47:18.2 (referred to as J1230.6+3247A),
and we consider this to be the most likely near-infrared counterpart.

{\bf J1231.2+2538} Compact radio structure with a faint \kband ID at its
centre.

{\bf J1231.3+3724} A compact radio sources with a faint \kband ID at its
centre.

{\bf J1231.5+3236} This source has a very extended radio morphology with
a \kband ID at its centre. There also seems to be an overdensity of
radio sources close to the radio source, possibly
indicative of a cluster.

{\bf J1232.2+2814} Another compact radio source with a faint \kband ID at
its centre.

{\bf J1232.6+3157} A small but extended radio morphology with two bright
\kband counterparts towards each of the extended components. This could be a
superposition of two sources or one radio source with two plausible
IDs, or alternatively it could be lensed. Spectroscopy will be needed to distinguish between these
possibilities. The \kband magnitudes of both the northern object
(J1232.6+3157N) and southern object (J1232.6+3157S) are given in
tables~\ref{tab:mags} and \ref{tab:summary}. Given the similarity in
\kband magnitude, the distribution in \kband magnitude of this sample
is not affected significantly by choosing one counterpart over the
other for this radio source.

{\bf J1234.3+2605} A double-lobed radio source with a plausible \kband
ID towards the eastern lobe. There is also a fainter \kband source at
the centre of the eastern lobe at $\alpha=$12:34:24.12,
$\delta=$26:05:51.4, both are within 1.5~arcsec of the centre of the
radio lobe and either could be the ID. We assume for this paper that
the brighter source is the ID as it lies along the radio axis.

{\bf J1238.2+2613} A compact radio sources with a \kband ID at its
centre.

{\bf J1238.8+3559} A double-lobed radio source with a bright \kband source
at the centre of the radio
structure, which we take to be the ID. There are also two further
\kband sources stretching along the western lobe, which could be due
to jet-induced star formation, although spectroscopy will be needed to
confirm that these objects are at the same redshift as the radio source.

{\bf J1243.7+2830} A compact radio structure with a \kband ID at its
centre. There is also another \kband source of similar magnitude that may
be interacting with this central galaxy $\sim 1.3$~arcsec to the north-east.

{\bf J1245.9+3320} A slightly extended radio source with a faint \kband
ID at its centre.

{\bf J1246.4+2516} A compact radio source with a faint \kband ID in the
centre.

{\bf J1248.2+2747} This source has a slightly extended radio morphology
with a fairly bright \kband ID close ($\sim 1$~arcsec away) to the
centre.

{\bf J1249.0+3615} This is an extended radio source with a bright \kband
ID at the centre of the brightest point in the radio emission, which we
take to be the core. This galaxy has a very close possible companion
$\sim 1$~arcsec to the north with which it may be interacting. The
74~MHz centroid for this source is $\sim 12$~arcsec to the north of the
core position quoted here. This may be because the 74~MHz emission
arises from a more extended, optically thin lobe.

{\bf J1249.7+3408} A compact radio source with a very faint \kband ID at the
centre. There are also three fainter sources in the
immediate vicinity ($< 6$~arcsec away) of the ID which may be
associated with an overdensity of galaxies at high redshift (the
\kband magnitude of this source means it is likely to be at $z > 4$).

{\bf J1250.4+2941} This source is a highly extended double-lobed radio
source in the East-West
direction. There is a bright ID at the centre of the source with a
line of three fainter sources aligned along the western lobe. This
 is highly unlikely to be a chance alignment, therefore these
three sources may be due to gas compression by the jet causing
increased star formation, or could be due to a preferred orientation
of galaxy overdensities with respect to the jet axis. Spectroscopy
will be needed to confirm whether these are actually at the same
redshift as the radio source.

{\bf J1252.7+2207} This is a fairly compact source with a \kband ID near
its centre. The astrometry on this image is only good to $\sim
1.5$~arcsec, therefore it is very plausible that the ID is actually at
the centre of the radio centroid.

{\bf J1253.4+2703} This source has a very unusual morphology in the
radio, and in Cohen et al. (2004) we suggested that it could be a
cluster relic. There is a bright \kband source near the centre of the
radio emission but spectroscopy will be needed to confirm whether this
is associated with the radio emission.

{\bf J1253.6+2509} This is a double-lobed radio source with a \kband ID
centred on the northern most extent of the southern lobe, along the
axis of the radio emission.

{\bf J1256.9+2811} This is a compact source with a very faint \kband
ID consistent with being at its centre.

\begin{figure*}


{\caption{\label{fig:kband_images}  \kband images (greyscale) of the
    26 USS sources from our 74~MHz sample. The images have been
    smoothed for presentation purposes. The black contours represent
    the radio maps from the 1.4~GHz FIRST survey. 
The contour levels are incremented by 10 per cent of the peak flux density
unless stated otherwise. are as follows: J1225.0+2146, Peak flux density=23.05~mJy/beam, lowest
contour is 10 per cent of the peak flux density. J1226.3+2418, Peak flux density=2.29~~mJy/beam, lowest contour is 35 per
cent of the peak. J1228.9+3114, Peak
flux density=1.88~mJy/beam, lowest contour is 30 per cent of the peak. J1229.1+3040, Peak
flux density=6.32~mJy/beam, lowest contour is 20 per cent of the peak. J1229.9+3712, Peak
flux density=5.54~mJy/beam, lowest contour is 20 per cent of the peak. J1230.2+2326, Peak
flux density=0.92~mJy/beam, lowest contour is 55 per cent of the peak with increments of 8 per cent.    
}}
\end{figure*}

\addtocounter{figure}{-1}

\begin{figure*}
{\caption{\label{fig:ov2} cont. The contour levels are incremented by 10 per cent of the peak flux density
unless stated otherwise. are as follows: J1230.6+3247, Peak flux density=0.59~mJy/beam, lowest
contour is 60 per cent of the peak flux density with increments of 8 per cent. J1231.2+2538, Peak flux density=15.55~mJy/beam, lowest contour is 10 per
cent of the peak. J1231.3+3724, Peak
flux density=2.75~mJy/beam, lowest contour is 30 per cent of the peak. J1231.5+3236, Peak
flux density=0.50~mJy/beam, lowest contour is 70 per cent of the peak  with increments of 5 per cent. J1232.2+2814, Peak
flux density=5.26~mJy/beam, lowest contour is 20 per cent of the peak. J1232.6+3157, Peak
flux density=1.94~mJy/beam, lowest contour is 40 per cent of the peak.}}
\end{figure*}

\addtocounter{figure}{-1}

\begin{figure*}

{\caption{\label{fig:ov2} cont.  The contour levels are incremented by
10 per cent of the peak flux density unless stated otherwise. are as follows:
J1234.3+2605, Peak flux density=1.73~mJy/beam, lowest contour is 40 per cent of the peak
flux density. J1238.2+2613, Peak flux density=15.68~mJy/beam, lowest contour is 10 per cent of the
peak. J1238.8+3559, Peak flux density=4.16~mJy/beam, lowest contour is 20 per cent of the
peak. J1243.7+2830, Peak flux density=0.98~mJy/beam, lowest contour is 50 per cent of the
peak with increments of 8 per cent. J1245.9+3320, Peak flux density=4.45~mJy/beam, lowest
contour is 20 per cent of the peak. J1246.4+2516, Peak flux density=4.25~mJy/beam, lowest
contour is 20 per cent of the peak.}}
\end{figure*}

\addtocounter{figure}{-1}

\begin{figure*}
{\caption{\label{fig:ov2} cont. The contour levels are incremented by
10 per cent of the peak flux density unless stated otherwise. are as follows:
J1248.2+2747, Peak flux density=4.34~mJy/beam, lowest contour is 20 per cent of the peak
flux density. J1249.0+3615, Peak flux density=2.05~mJy/beam, lowest contour is 35 per cent of the
peak. J1249.7+3408, Peak flux density=34.45~mJy/beam, lowest contour is 10 per cent of the
peak. J1250.4+2941, Peak flux density=1.59~mJy/beam, lowest contour is 40 per cent of the
peak. J1252.7+2207, Peak flux density=6.66~mJy/beam, lowest
contour is 10 per cent of the peak. J1253.4+2703, Peak flux density=0.81~mJy/beam, lowest
contour is 65 per cent of the peak with increments of 8 per cent.}}
\end{figure*}

\addtocounter{figure}{-1}

\begin{figure*}
{\caption{\label{fig:ov2} cont.  The contour levels are incremented by
10 per cent of the peak flux density unless stated otherwise. are as follows:
J1253.6+2509, Peak flux density=0.55~mJy/beam, lowest contour is 70 per cent of the peak
flux density with increments of 6 per cent. J1256.9+2509, Peak flux density=6.11~mJy/beam, lowest contour is 15 per cent of the
peak.}}
\end{figure*}

\begin{figure}
{\caption{\label{fig:dssimage} POSS-II $R-$band image of
    J1229.9+3712. The contours represent the FIRST source which is
    almost certainly a large source extending to $\sim 3$~arcmin. The
    \USS source is situated at the position of the western lobe. The
    peak flux density of the radio emission is 18.60~mJy/beam, and the lowest
    contour is at 8 per cent this value, with increments of 15 per cent. }}
\end{figure}

\begin{table*}
{\caption {\label{tab:summary}Summary of the key observational results for the \USS
  sample. The quoted RA and Dec are for the positions of the \kband
  IDs and all are J2000. $\dagger$ is a large source where the 74~MHz
  ID is associated with an extended lobe (Figure~\ref{fig:dssimage}); for
  $\ddagger$ the quoted magnitude relates to the \kband ID at J1230.6+3247A; the \kband magnitudes marked with a
  $^{\star}$ are measured using a 5~arcsec diameter aperture,
  $^{\star\star}$ is measured in a 3~arcsec apertures, given due to the
  proximity of a \kband source to the ID. The quoted lower limit for
  J1229.1+3040 is the 3$\sigma$ limit in an 8~arcsec aperture. For $^{\#}$ the quoted
magnitudes are for the bright galaxy to the north east, the actual ID
may not be visible in this image (see notes on this source). }}
\begin{center}
\begin{tabular}{lrccrl}
\hline\hline \mc{1}{c}{Source} & \mc{1}{c}{RA} & \mc{1}{c}{Dec} &
\mc{1}{c}{74~MHz flux density / Jy} &
\mc{1}{c}{$\alpha_{74}^{1400}$}& \mc{1}{c}{$K$ mag (8~arcsec)} \\
\hline\hline 
J1225.0$+$2146 & 12 25 02.40 & +21 46 52.6  & 2.213 & -1.25 & 19.54 $\pm$ 0.32 \\ 
J1226.3$+$2418 & 12 26 21.11 &+24 18 50.3 & 0.943 & -1.36 &  16.608 $\pm$ 0.080 \\ 
J1228.9$+$3114$^{\#}$ & 12 28 59.56 &+31 15 01.1 & 0.608 & -1.36 &  16.073 $\pm$ 0.069 \\
J1229.1$+$3040 & 12 29 07.75 &+30 40 41.2 & 0.356 & -1.27 & $> 19.7$ \\ 
J1229.9$+$3712$\dagger$ & 12 30 02.98& +37 12 46.1 & 0.635 & -1.25 &  \\ 
J1230.2$+$2326 & 12 30 14.31& +23 26 14.5 & 0.968 & -1.63 & 13.641 $\pm$ 0.019\\
J1230.6$+$3247A$\ddagger$ & 12 30 37.96 &+32 47 18.2 & 0.188 & $< -1.4$ & 19.655 $\pm$ 0.502 \\ 
J1231.2$+$2538 & 12 31 15.00 &+25 38 28.0 & 1.064 & -1.20 & 18.759 $\pm$ 0.225  \\ 
J1231.3$+$3724 & 12 31 20.71 &+37 24 14.7 & 0.477 & -1.51 & 17.814  $\pm$ 0.177 \\ 
J1231.5$+$3236 & 12 31 32.78 &+32 36 28.1 & 0.311 & -1.52 & 17.276 $\pm$ 0.119$^{\star}$ \\ 
J1232.2$+$2814 & 12 32 13.71 &+28 14 33.8 & 0.327 & -1.36 & 19.363 $\pm$ 0.292$^{\star}$\\ 
J1232.6$+$3157N & 12 32 38.13& +31 57 56.2 & 0.359 & -1.35 & 17.151 $\pm$ 0.097\\
J1232.6$+$3157S & 12 32 38.25& +31 57 50.7 & 0.359 & -1.35 & 16.726 $\pm$ 0.078\\
J1234.3$+$2605 & 12 34 23.95 &+26 05 49.7 & 0.402 & -1.61 & 17.786 $\pm$ 0.176 \\ 
J1238.2$+$2613 & 12 38 12.42 &+26 13 44.0 & 0.728 & -1.30 & 17.662 $\pm$ 0.162 \\ 
J1238.8$+$3559 & 12 38 49.44 &+35 59 21.4 & 0.634 & -1.33 & 16.522$\pm$  0.087\\ 
J1243.7$+$2830 & 12 43 42.80 &+28 30 55.5 & 0.712 & -1.73 & 16.717 $\pm$ 0.096 \\ 
J1245.9$+$3320 & 12 45 54.30 &+33 20 33.0 & 1.140 & -1.40 & 17.695 $\pm$ 0.151\\ 
J1246.4$+$2516 & 12 46 24.55 &+25 16 32.4 & 0.347 & -1.29 & 18.516 $\pm$ 0.223 \\
J1248.2$+$2747 & 12 48 13.62 &+27 47 22.8 & 0.624 & -1.31 & 17.413 $\pm$ 0.143\\
J1249.0$+$3615 & 12 49 01.46 &+36 15 34.4 & 0.440 & -1.20 & 15.500 $\pm$ 0.047 \\ 
J1249.7$+$3408 & 12 49 43.43 &+34 08 08.5 & 1.383 & -1.23 & 19.483 $\pm$ 0.340\\ 
J1250.4$+$2941 & 12 50 28.45 &+29 41 44.2 & 0.395 & -1.35 & 16.815 $\pm$ 0.092$^{\star\star}$\\ 
J1252.7$+$2207 & 12 52 44.04 &+22 07 01.5 & 0.940 & -1.22 &  17.447 $\pm$ 0.145 \\ 
J1253.4$+$2703 & 12 53 27.96 &+27 03 49.9 & 0.838 & -1.30 & 16.511 $\pm$ 0.087\\ 
J1253.6$+$2509 & 12 53 39.78 &+25 09 50.5 & 0.472 & $< -1.7$ & 18.503 $\pm$ 0.322\\
J1256.9$+$2811 & 12 56 58.15 &+28 11 09.7 & 0.524 & -1.32 & 18.885 $\pm$ 0.257\\
\hline\hline
\end{tabular}
\end{center}
\end{table*}

\section{Comparison with other samples}\label{sec:analysis} 

In this section we compare the distributions of $K-$band magnitude of
the USS sample with the distribution of magnitudes from a similar
survey selected at 151~MHz and also with a steep-spectrum filtering
criteria, namely 6C* (Jarvis et al. 2001a; 2001b). The \kband magnitudes can be used to estimate the redshift
of a source given the observed tightness of the $K-z$ relation
(e.g. Jarvis et al. 2001b; Willott et al. 2003).
Figure ~\ref{fig:histogram} shows the histogram of the \kband magnitudes
for the USS sample and the 6C* sample. 

A Kolmogorov-Smirnov test shows that the two datasets are
consistent with being drawn from the same underlying distribution, with the probability that they are
drawn from different distributions significant at $<
2\sigma$. However, figure~\ref{fig:histogram} does show that there are
more bright near-infrared sources in the 74-MHz-selected USS
sample. This is predominantly due to the fact that the 6C* sample has
an angular size limit of 15~arcsec, which may effectively filter out
closer objects which are not excluded from the 74 MHz sample. Indeed,
of those sources with $K < 17$, four of them have angular sizes $>
15$~arcsec. Removing these sources from the distribution results in
the distribution in the \kband magnitudes for the \USS and 6C* samples
being completely indistinguishable, with the probability that they
were drawn from the same underlying distribution $> 99$ per cent ($> 5\sigma$).

Given that the 6C* sample has a median redshift of $z \sim 1.9$
(Jarvis et al. 2001a), then we would expect that the USS sample
discussed in this paper would have a similar redshift distribution,
based on their \kband magnitudes. This is a much higher median
redshift than unfiltered, complete surveys of similar flux-density
limit, such as 7CRS where the median redshift is $z \sim 1.1$ (Willott et al. 2002). Therefore, we are confident
that the USS technique is still viable at 74~MHz, and as such we
envisage that this sample will contain some of the highest redshift
radio galaxies found to date.


\begin{figure}
{\caption{\label{fig:histogram} Histogram of the \kband magnitudes of
    the \USS sample (solid line) and the 151~MHz selected 6C* sample
    (dashed line).}}
\end{figure}

\section{Summary}\label{sec:conclusions}

We have completed \kband imaging of a complete sample of
\USS sources selected from a region spanning 165~deg$^{2}$. This
sample has been filtered using the steep-spectral index criterion to
target the highest-redshift radio galaxies. We find that the
distribution in the \kband magnitudes is similar that from a 
similar survey (6C*) undertaken at the slightly higher frequency of
151~MHz.

However, the \USS sample is relatively overabundant in brighter
sources when compared with the 6C* sample. This is predominantly due
to the angular size cutoff used for the 6C* sample which is not used
for the \USS sample here. Using this selection criteria on the \USS
sample would remove this bright near-infrared magnitude tail and the
probability that the two samples are drawn from the same underlying
distribution is significant at $>99$ per cent.

The faint tail of the \kband magnitude distribution
implies that we are still sensitive to the highest-redshift radio
galaxies, with our faintest source having $K > 20.7$ (3~arcsec
aperture), suggesting a redshift of $z > 5$ if we extrapolate the $K-z$ relation
for radio galaxies to $z > 5$.

The next stage in this project will be to gain spectroscopic redshifts
of all of the sources in the sample, which will enable us to find some
of the highest redshift radio galaxies and also constrain the
evolution in the comoving space density of the radio source population.

\section*{ACKNOWLEDGEMENTS} 
MJJ acknowledges the support of a PPARC PDRA. MJC acknowlegdes the support from the Portuguese Funda\c{c}\~{a}o
para a Ci\^{e}ncia e a Tecnologia.
The United Kingdom Infrared
Telescope is operated by the Joint Astronomy Centre on behalf of the
U.K. Particle Physics and Astronomy Research Council. Basic research in radio astronomy at the Naval
Research Laboratory is supported by the Office of Naval Research. The
National Radio Astronomy Observatory is a facility of the National Science
Foundation operated under cooperative agreement by Associated Universities, 
Inc.
The Digitized Sky Surveys were produced at the Space Telescope Science Institute under U.S. Government grant NAG W-2166. The images of these surveys are based on photographic data obtained using the Oschin Schmidt Telescope on Palomar Mountain and the UK Schmidt Telescope. The plates were processed into the present compressed digital form with the permission of these institutions. The Second Palomar Observatory Sky Survey (POSS-II) was made by the California Institute of Technology with funds from the National Science Foundation, the National Geographic Society, the Sloan Foundation, the Samuel Oschin Foundation, and the Eastman Kodak Corporation.

{}



\end{document}